\documentstyle[prb,aps,epsf,multicol]{revtex}

\begin{document}        
\draft
\title{Nonequilibrium dislocation dynamics and instability of 
driven vortex lattices in two dimensions}
\author{Igor S. Aranson,$^1$ Stefan Scheidl,$^{1,2}$ 
  and Valerii M. Vinokur$^1$}
\address{$^1$Materials
  Science Division, Argonne National Laboratory, Argonne, IL 60439\\
$^2$Institut f\"ur Theoretische Physik, Universit\"at zu
  K\"oln, Z\"ulpicher Str. 77, D-50937 K\"oln, Germany}

\date{\today}
\maketitle
\begin{abstract}
  We consider dislocations in a vortex lattice that is driven in a
  two-dimensional superconductor with random impurities.  The
  structure and dynamics of dislocations is studied in this genuine
  nonequilibrium situation on the basis of a coarse-grained equation
  of motion for the displacement field.  The presence of dislocations
  leads to a characteristic anisotropic distortion of the vortex
  density that is controlled by a Kardar-Parisi-Zhang nonlinearity in
  the coarse-grained equation of motion.  This nonlinearity also
  implies a screening of the interaction between dislocations and
  thereby an instability of the vortex lattice to the proliferation of
  free dislocations.
\end{abstract}


\begin{multicols}{2}

\section{Introduction}

Transport of periodic media such as vortex lattices in superconductors
and charge-density waves through random environments plays a
paradigmatic role in condensed matter physics.  While the pinning
dominated low-drive regime exhibiting glassy features has long been a
subject of extensive research, the nontrivial properties of the
high-velocity regime were recognized only recently.  The prediction of
a disorder induced nonequilibrium phase transition \cite{KV94} from
plastic to coherent motion of the vortex lattice upon increasing
drive, triggered extensive studies of rapidly driven disordered
lattices and have attracted much recent interest in theory,
\cite{KV94,BF95,mg,SV97-hv,BMR97-hv,GL97-hv} experiments,
\cite{BH93,Hel+96,Kee+97} and simulations.
\cite{MSZ96,FMM96,Ryu+96,DGB97,SJ97,OSZ+98}

The observed dynamic transition can be described qualitatively as
dynamic ``melting'' in analogy to the equilibrium melting transition,
where disorder induces an effective ``shaking''
temperature.\cite{KV94} However, this analogy cannot be extended to
the important questions regarding the role of disorder in structural
transformations of periodic structures.  The intrinsic nonequilibrium
nature of the driven state renders the analysis of the structural
transition specifically challenging. Even the {\it equilibrium}
melting is described theoretically in the two-dimensional case only.
Within the Kosterlitz-Thouless (KT) theory\cite{melt2d} melting is
mediated by the unbinding of dislocation pairs. The underlying melting
mechanisms are by far less understood in higher dimensions.  Yet the
issue of the stability of the ordered phase with respect to the
formation of topological defects was identified as a key issue for the
structural transitions, and substantial progress was achieved in
understanding the role and contribution of disorder in the static
case.\cite{kier} Whereas in equilibrium the criteria for the stability
of topological order follow from comparing relevant energy scales, the
analysis of defect nucleation in the {\em nonequilibrium} situation is
more subtle since it can no longer rely on energy balance
considerations.  In this paper we undertake a study of the stability
of the topological order of the dynamic state focusing specifically on
the two-dimensional vortex lattice driven through a disordered
environment.  We describe an intrinsic nonequilibrium mechanism giving
rise to the proliferation of topological defects and therefore the
instability of the driven vortex lattice.

The outline of this paper is as follows: in Sec. \ref{sec.mod} we
specify the coarse-grained equation of motion for the displacement
field. In Sec. \ref{sec.lin} we study the anisotropic and subdiffusive
nature of the dynamics of a single dislocation neglecting the effect
of the Kardar-Parisi-Zhang (KPZ) nonlinearity, which is discussed
further in Sec.  \ref{sec.nolin}.  There we show that the KPZ term
leads to a ``spiral'' structure of dislocations, screening their
long-range interaction, and recovering a normal linear diffusion of
defects. Our results are summarized and discussed in Sec.
\ref{sec.disc}. Some technical aspects are deferred to the Appendix.

\section{model}
\label{sec.mod}

We examine a dilute system of ``test'' dislocations embedded into the
elastic medium of the two-dimensional vortex lattice.  The vortices
are driven along one of the principal lattice directions, the $x$
axis.  The dynamics can be formulated in terms of a Langevin equation
of motion for the displacement $u$ of vortices from their perfect
lattice positions that move with average velocity $v$.
\cite{note.notation} The motion is governed by a competition of
elastic interactions between vortices, thermal noise, and pining
forces.  Pinning is described by a potential $V$ with local
correlation $\overline {V({\bf R}) V({\bf 0})} = \Delta_0 \delta({\bf
  R})$, where the $\delta$ function is supposed to have a width of the
order of the superconducting coherence length $\xi$.

Since the unbinding of dislocations is controlled by their interaction
on large scales it is legitimate to use a coarse-grained description
of the vortex lattice.  It was derived
recently\cite{BF95,SV97-hv,BMR97-hv,GL97-hv} that on large spatial
scales the equation of motion assumes the form
\begin{eqnarray}
\label{eqmo} 
   \eta \dot u &=& c {\bbox \nabla}^2 u + \delta F +\zeta 
    +\chi \nabla_x u + \frac \lambda 2 {(\bbox \nabla u)}^2 
    + f ({\bf r}+{\bf v} t ).
\end{eqnarray}
For our purposes it is sufficient to retain only the component of the
displacement field parallel to the velocity.  Although the other
components also experience fluctuations on large scales, they can only
further increase the instability of the lattice, which we find below
to occur even in the absence of transverse displacements.
Equation~(\ref{eqmo}) is written in the frame moving with the average
velocity of the vortices, where each vortex has a vanishing average
displacement around its average position ${\bf r}$.  $\eta$ is the
vortex friction coefficient.  For simplicity we consider the elastic
interaction as uniform, i.e., we use only one elastic constant $c$,
ignoring a distinction between shear and compression moduli and
additional anisotropic corrections obtained from coarse graining.
$\zeta$ is a thermal noise with temperature $T$.  Although $\eta$, $c$
and $T$ are renormalized under coarse graining, the corrections are
small in comparison to the original values for sufficiently large
drift velocities.  The stress coefficient $\chi$ and the
Kardar-Parisi-Zhang (KPZ) nonlinearity $\lambda$,\cite{KPZ86} which
are absent in the bare equation of motion, are generated under coarse
graining. The pinning force $f$, which is simply a derivative of the
pinning potential in the bare equation of motion, acquires a
random-force character under renormalization.  In principle, it
depends on the displacement as $f({\bf r} + {\bf v}t + {\bf u})$.
However, this dependence can be neglected after the coarse-graining
has been performed and $f$ has acquired random-force character.  In
the limit of large drift velocities these parameters take the
following values:\cite{SV97-hv}
\begin{mathletters}
\label{param}
\begin{eqnarray}
\chi &\approx& \frac{\Delta_0 c^2}{\xi^3 a^3 \eta^3 v^3},
\\
\lambda &\approx& \frac{\Delta_0 c^2}{\xi^4 a^2 \eta^3 v^3},
\label{lambda} 
\\
\delta F &\approx& \frac {\Delta_0}{\xi^4 \eta v},
\\
\overline{f({\bf R})f({\bf 0})} &\approx& 
\frac{\Delta_0^2}{\xi^6 \eta^2 v^2} \delta({\bf R}),
\\
\langle \zeta({\bf r},t) \zeta({\bf 0},0) \rangle &=&
2 \eta T \delta({\bf r}) \delta(t).
\end{eqnarray}
\end{mathletters}
$f$ and $\zeta$ are assumed to be Gaussian distributed with zero mean.
$a$ is the vortex lattice spacing.  The actual driving force $F=\eta v
+ \delta F$ required to achieve the prescribed velocity $v$ is
determined from the consistency condition that the average
displacement must vanish.

The linear stress term in Eq.~(\ref{eqmo}) can actually be eliminated
by a transformation of the displacement $\tilde u (\tilde {\bf r},t)=
u \left(\tilde {\bf r}-(\chi / \eta) t \hat {\bf x},t \right)$ to a
new frame moving with velocity $\tilde v=v-\chi/\eta$ in the
laboratory frame.\cite{note.trafo} The coordinates are related by
$\tilde {\bf r}={\bf r} + (t \chi / \eta) \hat {\bf x}$, where $\hat
{\bf x}$ denotes the unit vector along the velocity direction. The
following analysis is based on the transformed equation.  For
simplicity of notation we subsequently drop the tilde identifying
transformed quantities.

\section{Linear problem ($\lambda=0$)}
\label{sec.lin}

We first examine the model in the absence of the KPZ nonlinearity to
set the stage for its subsequent inclusion.  In the above model
dislocations are incorporated as a discontinuity in the displacement
field of amplitude $a$, the vortex spacing.  Due to the restriction of
our consideration to the displacement component parallel to the
velocity, the only possible orientations of Burgers vectors are
parallel or antiparallel to the velocity.

In order to derive the dynamic response of a dislocation to the
fluctuations of the elastic medium it is convenient to split the
displacement field into a topologically nontrivial (multi-valued) part
$u_0$ and a single-valued part $u_\sim$,
\begin{mathletters}
\begin{eqnarray}
u({\bf r},t)&=&u_0({\bf r}-{\bf r}_{\rm d}(t))+u_\sim({\bf r},t),
\\
u_0({\bf r})&\equiv& \frac a{2 \pi} \ \varphi({\bf r}).
\end{eqnarray}
\end{mathletters}
The dislocation position is ${\bf r}_{\rm d}(t)$ and $\varphi({\bf
  r})$ is the angle enclosed between the $x$ axis and ${\bf r}$.  The
problem at hand is then to derive an equation of motion for $u_\sim$
and ${\bf r}_{\rm d}$ from Eq.~(\ref{eqmo}).  To this end we consider
for the moment ${\bf r}_{\rm d}(t)$ as given and find
\begin{equation}
\label{eqmo.w}
\eta \dot u_\sim - c {\bbox \nabla}^2 u_\sim = 
\zeta + f ({\bf r}+{\bf v} t ) - \eta \dot {\bf r}_{\rm d} 
\cdot {\bbox \nabla} u_0({\bf r}-{\bf r}_{\rm d}).
\end{equation}
Since this equation is linear in $u_\sim$ one can easily calculate
$u_\sim$ for given $\zeta$, $f$, and dislocation trajectory ${\bf
  r}_{\rm d}$.  Eventually we determine the dislocation velocity $\dot
{\bf r}_{\rm d}$ in response to the forces $\zeta$ and $f$ from a
condition of local equilibrium for the dislocation core,
\begin{equation}
\label{no.grad}
{\bbox \nabla} u_\sim ({\bf r}_{\rm d}(t),t)=0.
\end{equation}
This equation is valid for ``slow'' changes of $u_\sim$ such that the
displacement singularity ``instantly'' moves to a point where it is no
longer subject to forces because its environment $u_\sim$ is locally
homogeneous. $u_\sim$ can be directly obtained from Eq.~(\ref{eqmo.w})
that has to be imposed with a short-scale cutoff of order $a$.

Supposing that the fluctuations of the elastic medium are weak and the
dislocation displacement is slow, the equation of motion for the
dislocation can be written in the form
\begin{equation}
\label{eqmo.disl}
-i \omega  \eta_{\rm d}(\omega) {\bf r}_{\rm d}(\omega)= 
{\bf f}_{\rm d}(\omega)
\end{equation}
 with a Peach-Koehler-like force
\begin{equation}
f_\alpha (t) = \epsilon_{\alpha \beta} \nabla_\beta  
u_\sim^{\rm ext} ({\bf r}_{\rm d}(t),t)
\end{equation}
($\epsilon$ is the totally antisymmetric tensor; see the Appendix for
some intermediate steps). Here $u_\sim^{\rm ext}$ is the contribution
to $u_\sim$ arising from the external forces $\zeta$ and $f$ at {\em
  zero} dislocation velocity.  The dislocation drag coefficient
\begin{mathletters}
\begin{eqnarray}
\eta_{\rm d}(\omega) &=& \frac a2 
\int_{\bf k} \frac 1 {-i \eta \omega + c k^2}
\\ 
&\approx& \frac{a \eta}{8 \pi c}
\left( \ln \frac{4 \pi c}{\eta |\omega| a^2}
+ i \arctan \frac{4 \pi c}{\eta |\omega| a^2} \right)
\ {\rm for} \ |\omega| \ll \frac{4 \pi c}{\eta a^2}
\end{eqnarray}
\end{mathletters}
diverges logarithmically for small frequencies because of the
long-ranged response of the displacement field to the dislocation
motion. This divergence with vanishing frequency is equivalent to a
divergence with increasing system size at zero frequency.  It is
characteristic for two dimensions and has been found previously for
dislocations in vortex lattices retaining both displacement
components\cite{Bra69} and also in pattern forming
systems.\cite{siggia,BPK88}

The force entering Eq.~(\ref{eqmo.disl}) has zero mean and
correlations
\begin{mathletters}
\begin{eqnarray}
\overline{\langle f_{{\rm d}\alpha}(\omega) 
f_{{\rm d}\beta}(\omega') \rangle} &=&
\delta(\omega+\omega') 
\Phi_{\alpha \beta}(\omega),
\\
\Phi_{\alpha \beta}(\omega)&=&
\int_{\bf k} \tilde k_\alpha \tilde k_\beta
\frac{2 \eta T + g \delta(\omega + {\bf k} \cdot {\bf v})}
{\eta^2 \omega^2 + c^2 k^4},
\end{eqnarray}
\end{mathletters}
where the disorder contribution is proportional to $g \equiv
{\Delta_0^2}/{\xi^6 \eta^2 v^2}$ and $\tilde k_\alpha \equiv
\sum_\beta \epsilon_{\alpha \beta} k_\beta$.  It is instructive to
compare the relative strength of thermal and pinning contributions
\begin{mathletters}
\begin{eqnarray}
\Phi_{\alpha \beta}^{\rm th}(\omega) &\approx& 
\delta_{\alpha \beta} \frac{\eta T}{8 \pi c^2}\ln
\left[1+\left( \frac{4 \pi c}{\eta \omega a^2} \right)^2 \right],
\\
\Phi_{yy}^{\rm pin}(\omega) &\approx& 
\frac{g}{6 \pi^2 \eta^2 v^3} \left|\frac{\eta \omega}{c}
\right|^{1/2}
\quad {\rm for} \quad |\omega| \ll \frac {\pi^2 c}{\eta a^2},
\\
\Phi_{xx}^{\rm pin}(\omega) &\approx& 
\frac{g}{4 \pi^2 c^2 v} \left|\frac{c}{\eta \omega}
\right|^{1/2}
\quad {\rm for} \quad |\omega| \ll \frac {\pi^2 c}{\eta a^2}.
\end{eqnarray}
\end{mathletters}
At low frequency the pinning contribution to the $y$ component is
negligible in comparison to the thermal contribution, $\Phi_{yy}^{\rm
  pin}(\omega) \ll \Phi_{yy}^{\rm th}(\omega)$.  However, the pinning
contribution to the $x$-component dominates over the thermal
contribution, $\Phi_{xx}^{\rm pin}(\omega) \gg \Phi_{xx}^{\rm
  th}(\omega)$.

Since ${\bf f}_{\rm d}$ vanishes on average, the dislocation has mean
zero velocity in the frame of Eq.~(\ref{eqmo.disl}), i.e., their
velocity is smaller than that of the vortices by $\chi/\eta \propto
v^{-3}$.  The diffusive behavior of dislocations is readily found from
Eq.~(\ref{eqmo.disl}).  In the absence of disorder dislocations
actually move {\em subdiffusively},
\begin{equation}
\label{diff.pure}
\langle [r_{{\rm d}\alpha}(t)- r_{{\rm d}\alpha}(0)]^2\rangle 
\approx \frac {32 T}{a^2 \eta} \frac {|t|}{\ln (4 \pi c |t|/\eta a^2)} 
\end{equation}
for $|t| \gg {\eta a^2}/{4 \pi c}$.  Despite of this peculiar dynamic
behavior the fluctuation-dissipation theorem holds for dislocations in
the absence of disorder, which implies that dislocation pairs
nevertheless unbind at the KT transition temperature.

The presence of disorder can significantly affect the diffusive
behavior of the dislocation.  For qualitative purposes we may still
use the linear response approach, where the forces acting on the
dislocation are evaluated at the undisplaced dislocation position.
This approach should provide a good approximation for a considerable
time interval because of the logarithmic divergence of the drag
coefficient.  From the above equations one finds that the motion is
{\em superdiffusive} along the $x$-direction,
\begin{equation}
\label{diff.dirt}
\overline{\langle [x_{\rm d}(t)- x_{\rm d}(0)]^2 \rangle} \approx 
\frac{64 g c^{1/2}}{3 \pi v \eta^{5/2} a^2} 
\frac {|t|^{3/2}}{\ln^2(4 \pi c |t|/\eta a^2)} 
\end{equation}
on large time scales $|t| \gg {\eta^3 v^2 T^2}/{c g^2}$, where the
pinning contribution dominates over the thermal contribution.  The
validity of Eq. (\ref{diff.dirt}) is limited by the fact that the
nonlinear terms in the equation of motion for ${\bf r}_{\rm d}$ have
been neglected. This could lead on largest time scales to a further
renormalization of the dislocation velocity $\tilde v$.  However, a
quantitative calculation of the dislocation velocity is beyond the
scope of the present paper, since it would require the inclusion of
additional contributions to the equation of motion, such as those due
to Peierls barriers.  However, despite of the possibility that Eq.
(\ref{diff.dirt}) may not capture the true large-scale behavior one
may conclude from a comparison of Eq. (\ref{diff.dirt}) to Eq.
(\ref{diff.pure}) that the shaking effects of disorder correspond on
large scales to an effectively infinite temperature.  This result
provides a first indication for a disorder-driven unbinding of
dislocations.

\section{Nonlinear problem  ($\lambda=0$)}
\label{sec.nolin}

So far the quadratic KPZ nonlinearity has been excluded from our
analysis and now we address the question how it modifies the above
findings.  For simplicity, we initially drop the random force term and
the thermal noise to examine the structure and dynamics of single
dislocations as well as the interaction between dislocations.  In this
case Eq. (\ref{eqmo}) is reduced to the Burgers equation
\begin{eqnarray}
\label{eqmo1}
\eta \dot u &=& c {\bbox \nabla}^2 u
+\frac \lambda 2 {(\bbox \nabla u )^2} + \delta F.
\end{eqnarray}
In analogy to the usual pinning problem we consider the vortex drift
velocity, which enters the last equation through $\lambda$, as
prescribed and determine the related force contribution $\delta F$
from the stationarity condition $\dot u=0$.  Topological defects in
Eq.~(\ref{eqmo1}) were studied recently in the context of
pattern-forming systems,\cite{BPK88,hagan,akw1,akw2,pn,ACT98} where
they constitute the source of ``spiral waves.'' Therefore it is
possible to carry over part of the previously achieved analysis to the
present context.

It is convenient to perform the well-known Hopf-Cole transformation
that leads to a linear equation for the function $W \equiv
\exp(\lambda u/2c)$,
\begin{eqnarray}
\label{eqmo2}
 \eta \dot W &=& c {\bbox \nabla}^2 W  -c  k_0^2 W .
\end{eqnarray}
Looking for a stationary solution, the constant
\begin{equation} 
k_0^2 = - \frac{ \lambda }{2 c^2 } \delta F 
\label{omega} 
\end{equation} 
will be determined from the condition that the solution is not
singular at the core (for the simplest choice $k_0=0$ the solution $W$
exhibits a non-physical $r^{-1}$ singularity at $r=0$).  Note, that
the stationarity of $u$ implies that $\lambda$ and $\delta F$ must
have opposite sign.

Equation (\ref{eqmo1}) possesses a solution of the form
\begin{equation} 
u({\bf r}) = \frac {a}{2 \pi} \varphi({\bf r}) + \mu(r) ,
\label{burg1} 
\end{equation} 
where $\mu(r)$ is a rotation-symmetric contribution induced by the KPZ
nonlinearity. This solution still represents a topological defect with
the characteristic multi-valuedness of the displacement $u(\varphi+2 \pi
m) = u(\varphi ) + a m$ (our explicit considerations apply to $m=1$
only, the generalization to integer $m$ is straightforward).  The
angular and radial dependences of $W$ factorize,
\begin{equation} 
\label{disl1} 
W_0(r,\varphi) = \exp [ (a \lambda / 4 \pi c) \varphi] \ w(r)  ,
\end{equation} 
and for a stationary solution of Eq. (\ref{eqmo2}) the radial
contribution $w(r)=\exp[\lambda \mu(r)/ 2 c ]$ has to satisfy a
modified Bessel equation
\begin{equation} 
\partial_r^2 w + \frac{1}{r}  \partial_r w+ 
\frac{\alpha^2}{r^2} w = k_0^2 w,
\label{bess} 
\end{equation} 
where $\alpha \equiv a \lambda /4 \pi c$.  Thus the solution to Eq.
(\ref{bess}) is a modified Bessel function $w(r) = K_{i \alpha} (k_0
r) $ with imaginary index.\cite{note.I} This solution has two
characteristic length scales separating three regions.  On large
length scales $r \gg L_s \equiv k_0^{-1}$ this solution decays
exponentially, $w(r) \propto \exp[-k_0 r]/r^{1/2}$. For $r \ll
k_0^{-1}$ and $\alpha \ll 1$ an expansion of the Bessel function
yields $w(r) \approx \sin[\alpha(\ln (2/k_0 r) -\gamma)]$ with Euler's
constant $\gamma$.  Thus $w(r)$ assumes a maximum at a second
characteristic scale
\begin{equation}
r_0 \equiv k_0^{-1} e^{-\pi /2 \alpha}.
\label{r_0}
\end{equation}
For $r \ll r_0$ the solution becomes strongly oscillating, which is
unphysical.  It is important to keep in mind that the equation of
motion for the displacement field is valid only on scales larger than
a cutoff scale $R_c$ of the order of the vortex spacing.  Therefore
the oscillatory behavior of $w$ at small scales is an artifact of Eq.
(\ref{eqmo1}) that does not account for the dislocation core
structure.  The above solution is physically meaningful only as long
as $\mu(r)$ or $w(r)$ depend monotonously on the distance, i.e., the
outermost maximum of the solution found above has to be identified
with the core radius.  Thus Eq. (\ref{r_0}) actually determines $k_0$,
or equivalently the screening length $L_s$ as a function of
$\alpha$,\cite{note.regular}
\begin{mathletters}
\label{sl}
\begin{eqnarray} 
L_s &=& k_0^{-1}
\sim  R_c \ e^{\pi/ 2 \alpha}  \sim a \ e^{(v/v_0)^3} ,
\\
v_0^3 &=& \frac {\Delta_0 c}{2 \pi^2 \xi^4 a^3 \eta^3} .
\end{eqnarray}
\end{mathletters} 
For small $\alpha $, i.e., large velocity $v$ of the vortex lattice,
$L_s$ is exponentially large.  Since for $r \to \infty$ the solution
$w \sim \exp[ - k_0 r]/ r^{1/2}$ has exponential asymptotics, the
displacement $u \approx (2c/\lambda) \ln w \approx - (2c/\lambda) k_0
r$ increases {\em proportional to the distance} from the dislocation
core.

The exponential decay of the function $K$ readily implies {\em
  exponential screening} of the interaction between several
dislocations at positions ${\bf r}_i$ with distances large compared to
$L_s$. For pattern forming systems, this screening was a subject of
intensive investigations.\cite{akw1,akw2,pn} Indeed, since Eq.
(\ref{eqmo2}) is linear in $W$, the multi-dislocation solution can be
approximated as a linear superposition of individual solutions,
\begin{equation} 
W({\bf r}) = \sum_i  W_0(| {\bf r - r_i}| ) + \tilde w,
\label{dislm}  
\end{equation} 
where the correction $\tilde w$ is introduced in order to fix
``topological conditions'' imposed by the field $u$.  For a
well-separated ensemble of dislocations the overlap between the
individual contributions to $W$ is exponentially small. From the
velocity of a dislocation under the influence of other dislocations
one can actually obtain the dislocation interaction.  The resulting
interaction between two dislocations $i$ and $j$ decays as $\exp [ -2
k_0 | {\bf r_i - r_j}| ] $ for $k_0 | {\bf r_i - r_j}| \gg 1 $ (for
$k_0 | {\bf r_i - r_j}| < 1 $ there is a crossover to the usual
power-like behavior).\cite{akw2,note.interact}

\begin{figure}
\epsfxsize=0.9 \linewidth
\epsfbox{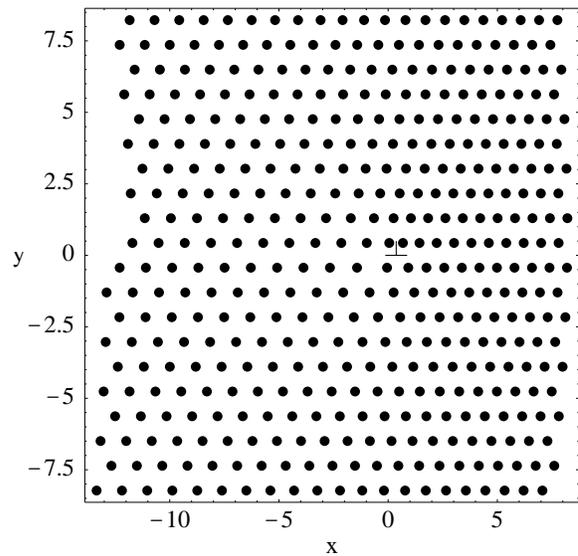}
\narrowtext
\caption{
  Schematic illustration of the structure of a single dislocation
  (marked by the symbol $\perp$) in the vortex lattice driven along
  the $x$ axis (lengths are in units of $a$). The density is
  increased/reduced in directions in front of/behind the dislocation.}
\label{fig_spiral}
\end{figure}

The structure of such a ``spiral'' dislocation is illustrated
schematically in Fig. \ref{fig_spiral}.  Note that the lattice is
compressed in front of the dislocation, whereas it is diluted behind
the dislocation.  This effect is independent of the orientation of the
Burgers vector parallel or antiparallel with the velocity.  Given the
fact that $\lambda$ is positive this effect can be easily understood
from Eq. (\ref{eqmo1}), if one considers the KPZ nonlinearity as
perturbation to the usual dislocation: the displacement strains are
largest close to the dislocation core.  Therefore the vortices close
to the core experience a force that drives them further in the
direction of the velocity until the elastic forces establish force
equilibrium in a configuration, where all vortices move with the same
velocity $v$.  

Due to the presence of the KPZ term in the equation of motion
(\ref{eqmo1}) the force required to achieve a vortex motion with
velocity $v$ is reduced: $\delta F = F-\eta v =-(2 c/\lambda)k_0^2<0$.
However, this contribution of the KPZ term, which is generated by
disorder, actually represents only a further correction to the
immediate pinning force (\ref{param}c), which is positive.  For large
$v$ the correction due to the KPZ term is negligible in comparison to
the pinning force, since the former contribution decays exponentially
for large $v$ whereas the latter decays only algebraically.
Considering both contributions consistently together, one finds $F
>\eta v$, i.e., even in the presence of the KPZ nonlinearity disorder
actually slows down the vortex motion.

\begin{figure}
\epsfxsize=1.0 \linewidth
\epsfbox{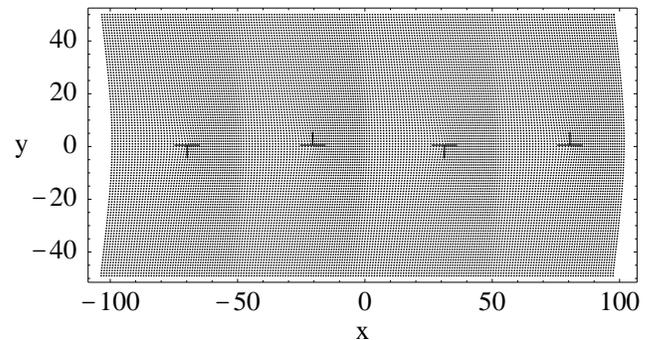}
\narrowtext
\caption{
  Schematic illustration of a dislocation pair (repeating two unit
  cells with periodic boundary conditions) in the driven vortex
  lattice.  The vortex density changes quite abruptly along pronounced
  shock wave fronts.  The vortex lattice is shown here in a square
  lattice to make the local rotations and compression of the unit cell
  graphically more transparent although in reality the vortex lattice
  is hexagonal.}
\label{fig_pair}
\end{figure}

The structure of the vortex lattice in the presence of a
dislocation/antidislocation {\em pair} is illustrated in Fig.
\ref{fig_pair}.  This structure was calculated numerically directly
from Eq. (\ref{eqmo1}) using a lattice domain with periodic boundary
conditions.  In Fig. \ref{fig_pair} two such domains are reproduced.
The inhomogeneity of the density is apparent and the density can
change along narrow boundaries that correspond to shock wave
fronts.\cite{KPZ86}

Let us now include the effect of weak thermal noise and disorder in
the dynamics of a single dislocation.  After the Hopf-Cole
transformation Eq. (\ref{eqmo}) assumes the form
\begin{eqnarray}
\label{eqmo3}
 \eta \dot W &=& c {\bbox \nabla}^2 W  -c  k_0^2 W + 
\frac{\lambda}{2 c} [\zeta + f({\bf r} + {\bf v} t) ] W .
\end{eqnarray}
To derive the resulting equation of motion for a dislocation we apply
a perturbation technique,\cite{akw2} where the force fields of thermal
noise and disorder are projected on the translation modes $W_{x,y}$ of
the unperturbed Eq.~(\ref{eqmo3}) (see also the Appendix).  Since the
translation mode is simply $W_{x,y} = \partial_{x,y} W_0$, the
equation of motion for the dislocation position ${\bf r_0}$ is of the
form
\begin{eqnarray} 
\eta_{\rm d} \partial_t {\bf r_0}  &=&
\frac{\lambda}{2 c} \int d^2 r'  
[ \zeta({\bf r}',t) + f({\bf r}' - {\bf r}_0(t)  + {\bf v} t)]
\nonumber \\ && \times
W_0 ({\bf  r ^\prime}) {\bbox \nabla}  W_0 ({\bf  r ^\prime})
\label{eqmo4} 
\end{eqnarray} 
with an effective drag coefficient $\eta_{\rm d}$ of the form
\begin{equation} 
\eta_{\rm d} = \frac{\eta}{2} \int d^2 r^\prime [ {\bbox \nabla}  
W_0 ({\bf r}^\prime)]^2.
\end{equation}
Because of the exponential localization of the function $W_0$, the
drag coefficient $\eta_{\rm d}$ is finite, in contrast to the linear
case, where it diverges logarithmically with the volume.  From Eq.
(\ref{eqmo4}) we obtain the mean squared displacement [abbreviating
$\zeta^{(n)} \equiv \zeta ({\bf r}^{(n)}, t^{(n)})$ and $f^{(n)}
\equiv f({\bf r}^{(n)} - {\bf r}_0(t^{(n)})  + {\bf v} t^{(n)})$]
\begin{eqnarray} 
\langle {\bf r}_0^2(t) \rangle 
&=&\frac{\lambda^2 }{4 c^2 \eta_{d}¥} 
\int_0^t  dt' \int _0^t  dt''  \int d^2 r' d^2 r''
\langle [\zeta' + f'] [\zeta'' + f'']\rangle
\nonumber \\ & & \times
W_0 ({\bf r}') {\bbox \nabla}  W_0({\bf r}'')  
W_0 ({\bf r}'') {\bbox \nabla}  W_0({\bf r}'')
\label{int} 
\end{eqnarray} 

Let us first consider only the effect of thermal noise $\zeta$.
Averaging Eq. (\ref{int}) and utilizing the exponential decay of $W$
at large $r$, we readily obtain a normal diffusive behavior,
\begin{equation} 
\langle {\bf r}_0^2(t) \rangle \sim t .
\end{equation} 
Thus, the exponential screening of the dislocation field due to the
KPZ nonlinearity is responsible for the transition from subdiffusion
[Eq.~(\ref{diff.pure})] to {\em normal} diffusion.

To include the effect of quenched disorder we have to perform an
additional disorder averaging of Eq. (\ref{int}).  Thereby we have to
assume that the position of the dislocation is not correlated with the
disorder, which is true at large enough drift velocity where the
dislocation moves with the vortices.  As in the case of purely thermal
noise we obtain normal diffusion of the dislocation due to the
exponential localization of $W$, in contrast to superdiffusion in the
linear case.

\section{Conclusions}
\label{sec.disc}

The most important implication of the exponential screening of the
dislocation interaction is that even arbitrarily weak thermal noise or
random force result in an {\em unbinding of dislocations} (see Ref.
\CITE{ACT98}).  This means that the corresponding Kosterlitz-Thouless
temperature is zero in this situation, and that the topological order
of the vortex lattice is always destroyed on largest scales.  While
previous arguments in favor of such an instability were based on
scaling arguments\cite{BF95} and numerical simulations\cite{SJ97} we
have presented here an intrinsic {\em nonequilibrium mechanism}.  This
instability mechanism is primarily based on the presence of the KPZ
nonlinearity in the equation of motion, which is generated by coarse
graining the equation of motion of a driven vortex lattice.
 
At this point we would like to recall that our analysis was based on
the simplified model with only one displacement component.
Fluctuations of the second component will probably lead to a
significant further reduction of the stability of the vortex lattice,
i.e., the true screening length could possibly be smaller than the
value for $L_s$ calculated above.

It is instructive to compare the dislocation screening length $L_s$ to
the crossover scale where the KPZ nonlinearity becomes relevant for
thermal fluctuations of the displacement field in the absence of
dislocations. The latter scale is $\xi_c \equiv \exp(8 \pi c^3/\eta T
\lambda^2)$,\cite{TNF90} i.e., it is also exponentially large for
small $\lambda$. However, the functional dependence of $\lambda$ is
different and $\L_s \ll \xi_c$ for small $\lambda$. [Note, that
$\xi_c$ is defined only for finite temperature while $L_s$ is
well-defined even for zero temperature.] Therefore, dislocations
become important for the structure of the system before it can show a
crossover to the strong-coupling behavior of the elastic system.

In a remote analogy the interaction of dislocation pairs in a driven
vortex lattice can be compared to the interaction of vortex pairs in a
superconducting film.  In the former case screening is induced by
disorder, which in the latter case is provided by magnetic fields.  In
the absence of screening both systems would perform a
Kosterlitz-Thouless transition. Screening suppresses this transition
as a large-scale phenomenon.  However, for large enough screening
lengths this crossover can still be well pronounced.  The screening
length of the vortex interaction in films is known to be macroscopic.
The screening length of dislocations $L_s$ as found above will also be
macroscopic for $v \gg v_0$ because of its exponential increase with
the vortex drift velocity.  It is therefore possible that the lattice
may be well ordered on experimentally relevant scales and that a
dynamic melting or freezing can still be found as a pronounced
crossover.

In the nonequilibrium mechanism for the dislocation unbinding examined
above the KPZ nonlinearity played a major role.  It is worthwhile to
point out that even if dislocations are artificially suppressed in a
purely elastic approach, this nonlinearity can induce dynamic
transitions between ``rough'' and ``flat'' sliding
phases.\cite{CBFM96} The above analysis demonstrates that even the
``flat'' sliding phase is actually penetrated by free dislocations on
length scales beyond $L_s$. Therefore one might suspect that their
presence modifies the nature of this roughnening transition or even
blurs the transition, turning it into a mere crossover. Again, this
crossover may yet be observable because $L_s$ grows exponentially for
large drift velocities.

To conclude, we have found an instability of the simplified
one-component model of a drifting vortex lattice to proliferation of
free defects.  As already mentioned above, the presence of a second
displacement component can only lead to a further increase of
fluctuations and to a reduction of the instability.  We have presented
an intrinsic nonequilibrium mechanism for such an instability based on
the presence of the KPZ nonlinearity in the coarse-grained equation of
motion for the displacement field of a vortex lattice driven in a
disordered environment.  The instability is due to an exponential
screening of the dislocation interaction on a scale $L_s$ that
increases exponentially with the drift velocity of the vortex lattice.

This work was supported by Argonne National Laboratory through the
U.S.  Department of Energy, BES-Materials Sciences, under contract No.
W-31-109-ENG-38 and by the NSF Science and Technology Center for
Superconductivity under contract No. DMR91-20000.  S. S. acknowledges
support from the Deutsche Forschungsgemeinschaft under Project No.
SFB341 and Grant No. SCHE/513/2-1.

\appendix

\section*{}

In this appendix we give some intermediate steps of the calculation
leading to equation of motion (\ref{eqmo.disl}) for the dislocation in
the linear medium. Furthermore we relate this calculation to a
projection of the force field onto the translation modes of the
dislocation, as used in Sec.~\ref{sec.nolin}.

For given ${\bf r}_{\rm d}$ Eq. (\ref{eqmo.w}) can be solved for
$u_\sim$. Equation (\ref{no.grad}) then implies
\begin{eqnarray*}
0 &=& \epsilon_{\alpha \beta} \nabla_\beta u_\sim({\bf r}_{\rm
d}(t),t) 
\\
&=& \epsilon_{\alpha \beta} \int d^2r' dt' \nabla_\beta 
G({\bf r}_{\rm d}(t)- {\bf r}',t-t')
\nonumber \\ &\times&
 [\zeta({\bf r}',t')+f({\bf r}'+{\bf v}t') 
- \eta \dot {\bf r}_{\rm d}(t') \cdot 
\nabla u_0({\bf r}'-{\bf r}_{\rm d}(t') )]
\end{eqnarray*}
where $G$ is the Greens function for $u_\sim$. In linear response,
$\zeta$, $f$, and hence $\dot {\bf r}_{\rm d}$ are small and one may
replace ${\bf r}_{\rm d}(t')$ by ${\bf r}_{\rm d}(t)$ in the previous
Equation. This equation has to be understood as
force balance between the contribution ${\bf f}_{\rm d}$ arising from
the external force fields $\zeta$ and $f$ and the friction force
$\eta_d \dot {\bf r}_{\rm d}$.  The separation of these contribution
leads to Eq.  (\ref{eqmo.disl}).

Let us now briefly discuss the connection between the approaches used
in Secs. \ref{sec.lin} and \ref{sec.nolin} to calculate the
dislocation dynamics.  The subdiffusive behavior of dislocations in
the linear system can be easily obtained from the equation of motion,
taking into account that in the linear system (\ref{eqmo.w}) the
translation mode is $u_{xy}= \partial_{x,y} u_0 ~\sim 1/r $.  Thus,
the effective mass diverges as $\ln r$.  For the mean-squared
displacement we obtain (dropping the disorder term)
\begin{eqnarray*}
\langle {\bf r}_0^2 (t) \rangle
&=& \frac{1 }{ \eta_{d}}
\int_0^t  \int _0^t  \int d^2 r' d^2 r'' dt' dt''
\nonumber \\ && \times
\langle \zeta' \zeta''  {\bbox \nabla}  u_0
({\bf  r ^ \prime}) {\bbox \nabla}  u_0
({\bf  r ^{\prime \prime}})
\rangle
\end{eqnarray*}
instead Eq. (\ref{int}). After averaging we obtain $\langle {\bf
  r}_0^2 (t) \rangle \sim t / \ln r \approx t/ \ln t $, which
coincides with Eq. (\ref{diff.pure}).

\end{multicols}

\begin{references}
\bibitem{note.notation} We follow the notations of Ref.
  \CITE{SV97-hv}.
\bibitem{note.trafo} Note that this transformation can be performed
  only neglecting the {\em discreteness} of the lattice. The
  discreteness can be important at temperatures small compared to the
  Peierls barriers for dislocation motion; see also S. Scheidl and
  V.M. Vinokur, \prb {\bf 56}, R8522 (1997).
\bibitem{note.I} The Bessel function $I$, which is also a solution to
  Eq. (\ref{bess}), cannot satisfy this condition at the core of the
  dislocation, see Ref. \CITE{hagan} for details.
\bibitem{note.regular} Equation (\ref{eqmo1}) can be regularized by a
  Ginzburg-Landau equation with complex coefficients. Within this
  scheme, where one can calculate the actual dislocation core
  structure and verify the matching procedure at the core radius and
  the resulting relation (\ref{sl}).\cite{hagan}
\end{references}
\end{document}